\documentclass[
  ,draft            % use draft while you are working on the paper
  ]
  {aipproc}

\layoutstyle{8x11single}

\begin{document}

\title{Simulation chain and signal classification for acoustic neutrino detection in seawater}

\classification{95.55.Vj, 95.75.Pq, 43.30.Cq, 43.30.Zk}
\keywords      {Acoustic particle detection, Neutrino detection, simulation, signal classification}

\author{D. Kie\ss{}ling}{
  address={Friedrich-Alexander-Universit\"at Erlangen-N\"urnberg, Erlangen Centre for Astroparticle Physics, Erwin-Rommel-Str. 1, 91058 Erlangen, Germany}
}

\author{G. Anton}{
  address={Friedrich-Alexander-Universit\"at Erlangen-N\"urnberg, Erlangen Centre for Astroparticle Physics, Erwin-Rommel-Str. 1, 91058 Erlangen, Germany}
}

\author{A. Enzenh\"ofer}{
  address={Friedrich-Alexander-Universit\"at Erlangen-N\"urnberg, Erlangen Centre for Astroparticle Physics, Erwin-Rommel-Str. 1, 91058 Erlangen, Germany}
}

\author{K. Graf}{
  address={Friedrich-Alexander-Universit\"at Erlangen-N\"urnberg, Erlangen Centre for Astroparticle Physics, Erwin-Rommel-Str. 1, 91058 Erlangen, Germany}
}

\author{J. H\"o\ss{}l}{
  address={Friedrich-Alexander-Universit\"at Erlangen-N\"urnberg, Erlangen Centre for Astroparticle Physics, Erwin-Rommel-Str. 1, 91058 Erlangen, Germany}
}

\author{U. Katz}{
  address={Friedrich-Alexander-Universit\"at Erlangen-N\"urnberg, Erlangen Centre for Astroparticle Physics, Erwin-Rommel-Str. 1, 91058 Erlangen, Germany}
}

\author{R. Lahmann}{
  address={Friedrich-Alexander-Universit\"at Erlangen-N\"urnberg, Erlangen Centre for Astroparticle Physics, Erwin-Rommel-Str. 1, 91058 Erlangen, Germany}
}

\author{M.~Neff}{
  address={Friedrich-Alexander-Universit\"at Erlangen-N\"urnberg, Erlangen Centre for Astroparticle Physics, Erwin-Rommel-Str. 1, 91058 Erlangen, Germany}
}

\author{C. Sieger}{
  address={Friedrich-Alexander-Universit\"at Erlangen-N\"urnberg, Erlangen Centre for Astroparticle Physics, Erwin-Rommel-Str. 1, 91058 Erlangen, Germany}
}

\begin{abstract}
Acoustic neutrino detection is a promising approach to extend the energy range of neutrino telescopes to energies beyond $10^{18}$\,eV. Currently operational and planned water-Cherenkov neutrino telescopes, most notably KM3NeT, include acoustic sensors in addition to the optical ones. These acoustic sensors could be used as instruments for acoustic detection, while their main purpose is the position calibration of the detection units. In this article, a Monte Carlo simulation chain for acoustic detectors will be presented, covering the initial interaction of the neutrino up to the signal classification of recorded events. The ambient and transient background in the simulation was implemented according to data recorded by the acoustic set-up AMADEUS inside the ANTARES detector. The effects of refraction on the neutrino signature in the detector are studied, and a classification of the recorded events is implemented. As bipolar waveforms similar to those of the expected neutrino signals are also emitted from other sound sources, additional features like the geometrical shape of the propagation have to be considered for the signal classification. This leads to a large improvement of the background suppression by almost two orders of magnitude, since a flat cylindrical ``pancake'' propagation pattern is a distinctive feature of neutrino signals. An overview of the simulation chain and the signal classification will be presented and preliminary studies of the performance of the classification will be discussed.
\end{abstract}

\maketitle

%%%%%%%%%%%%%%%%%%%%%%%%%%%%%%%%%%%%%%%%%%%%
%% MAINMATTER
%%%%%%%%%%%%%%%%%%%%%%%%%%%%%%%%%%%%%%%%%%%%

\section{Acoustic neutrino detection}
Acoustic neutrino detection is an approach to extend the energy range of water-Cherenkov neutrino telescopes up to the energies of the neutrino flux predicted by the GZK-effect~\cite{ref1}.The acoustic signal from these neutrinos is generated according to the thermoacoustic model~\cite{ref2} in the following way: the incindent neutrino interacts with the seawater and a hadronic cascade is formed. This causes a local energy deposition and heats the sourrounding matter. The water then expands, forming a pressure pulse emitted in a plane (often called the ``pancake''). The acoustic test system AMADEUS~\cite{ref3}, which is an integral part of the ANTARES detector~\cite{ref4}, has been operational since 2008. This set-up only comprises 36 hydrophones in 6 clusters, meaning that it is too small to discriminate the neutrino signal from background noise. However, the future water-Cherenkov telescope KM3NeT~\cite{ref5}, which is currently under construction, will also encompass an acoustic system for the position calibration of its optical modules. Therefore a sensor based on a piezo-ceramic element will be included in each of the 18 optical modules on every line. This system may also be used for acoustic particle detection, and it would be the largest acoustic detector to date. The possibilities of such a large scale acoustic detector are investigated. 
\section{Effects of refraction on simulations of acoustic detectors}
The speed of sound is not constant in seawater, as it varies with pressure, temperature and salinity of the water. Since these parameters all depend on the depth, the sound speed also has to change with depth. This leads to a variation of the index of refraction of the sea water, which causes the path of the sound waves to be bent. %
%See Fig.~\ref{profile} for a sound speed profile recorded at the ANTARES detector site. 
For the Mediterranean Sea, the sound is deflected upwards in the deep sea, since the speed of sound is increasing almost linearly with depth. %
It is an aspect for the simulation of an acoustic neutrino detector to have a correct implementation of the sound propagation. This means that a second order non-linear partial differential equation has to be solved, which can only be done numerically\footnote{An approximate analytical solution exists if one assumes a constant ratio of the speed of sound and its derivation regarding the depth.}. However, this procedure requires approximately 10\,000 times more floating point operations than required for a straight propagation, which would be the case for a constant speed of sound. Due to this huge difference in complexity, it is advisable to check whether the curved sound paths have an important effect on the simulations, or in other words if a constant sound speed in the whole ocean can be assumed. %
The deviation of the ray tracing model and the straight sound propagation is less than 20\,m per kilometer for the environmental parameters at the ANTARES detector site. Considering a typical water-based neutrino detector (e.g. KM3NeT), with a horizontal extension of approximately 500\,m and vertical spacings between sensors of 36\,m, this effect cannot be detected. The signatures in the detector generated by the different simulation methods cannot be distinguished since the sound wave is sent out in an approximately 20\,m thick plane. However, an error in the direction reconstruction is introduced. For a single event with a vertex position outside of the detector, the difference between the reconstructed incident direction and the simulated one can be as large as $5^{\circ}$. This error mainly depends on the distance of the interaction point to the center of the detector, so if only contained events are considered, this is not a problem. One of the main advantages of acoustic detection diminishes in this case as the long range of these signals of several kilometers is no longer exploited. 
Since the exact reconstruction of the vertex position and the incident direction are not relevant for a general estimation of the performance of the detector, and because the signature within the detector still looks like a plane, the ray tracing model will not be used for the following sections.
\section{Simulation chain for acoustic neutrino detection}
The simulation chain was implemented in a modular software framework written in C++ and python. The framework called SeaTray, is based on IceTray from the IceCube collaboration~\cite{ref6}. The framework itself has no knowledge of the actual physics. It just provides the means to split the code in logical blocks, called modules, and it also provides several necessary tools like geometry classes or random number generators. SeaTray is used for Monte Carlo (MC) simulations and the reconstruction of events. Due to its modular nature, the code written for one purpose can often be reused, or it can be easily replaced if for example a better reconstruction algorithmn becomes available. Another possibility provided by the modular structure of the framework is the ability to use simulation chains with different levels of detail.
\begin{figure}
  \includegraphics[height=.3\textheight,clip= true,trim= 0 0 0 30]{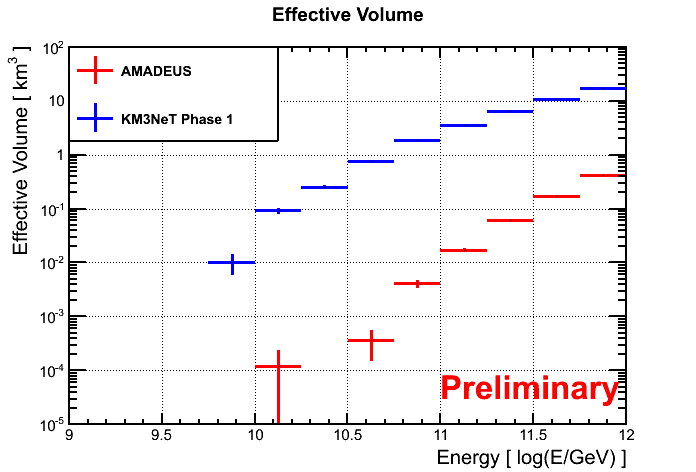}
  \caption{Comparison of the effective volume of AMADEUS set-up and the first phase of the KM3NeT detector, taking only a block of 24 lines into account. The simulations were performed with the simulation chain using parametrizations as described in the text.}
  \label{fig:volume}
\end{figure}
The most accurate results are gained with a full simulation chain, that uses a detailed MC simulation for every aspect of the detector. The first module generates a neutrino interaction with parameters in pre-defined ranges for energy, direction and position. Two other modules, the shower generator and the pulse generator create the acoustic signal from the hadronic cascade. The signal generation in the simulation was implemented according to a parametrization made by the ACoRNe collaboration~\cite{ref7}. The ambient noise is added using a model developed from the noise recorded by the AMADEUS set-up~\cite{ref8} and transient background signals can be added. The response of the sensors and electronics is simulated together with inherent noise of the system in another module. Finally, a filter simulation is available using a matched filter with a coincendence test, as it is implemented inside the AMADEUS data acquisition. These last three steps of the simulation chain are all dependent on the detector and were not yet adapted to the KM3NeT design. Producing enough events for high statistics with this simulation may require several months of computation, even using a lot of cores in parallel.
 
An alternative to the aforementioned approach is using parametrizations for most values, which reduces the required computing time and allows for a quick comparison of different detectors, even when the system response is not completely known yet. The simulation is altered after an event has been generated by the particle generator. The signal amplitude is then determined by another module using a parametrization, and it is checked against a fixed noise level. The individual hydrophone is triggered if the signal strength exceeds a certain signal to noise ratio.  %
See Fig.~\ref{fig:volume} for a plot of the simulated effective volumes of the AMADEUS set-up and a detector of the same size as the first phase of the KM3NeT detector. For each bin, 1\,000\,000 events were generated within a radius of 12\,km around the detector center. The noise level was assumed to be 15\,mPa (corresponding to the average noise level recorded within the AMADEUS set-up) and the minimum signal to noise ratio required for a hit was 2. A minimum of six sensors had to trigger a signal in coincidence to consider it as detected and it is assumed that the background is fully supressed. The results show an increase of the effective volume by almost two orders of magnitude, meaning that the acoustic system of the KM3NeT telescope had a lot of potential if it would be used for acoustic detection.

\section{Signal classification}
\begin{figure}
  \includegraphics[height=.3\textheight,clip= true,trim= 0 0 0 30]{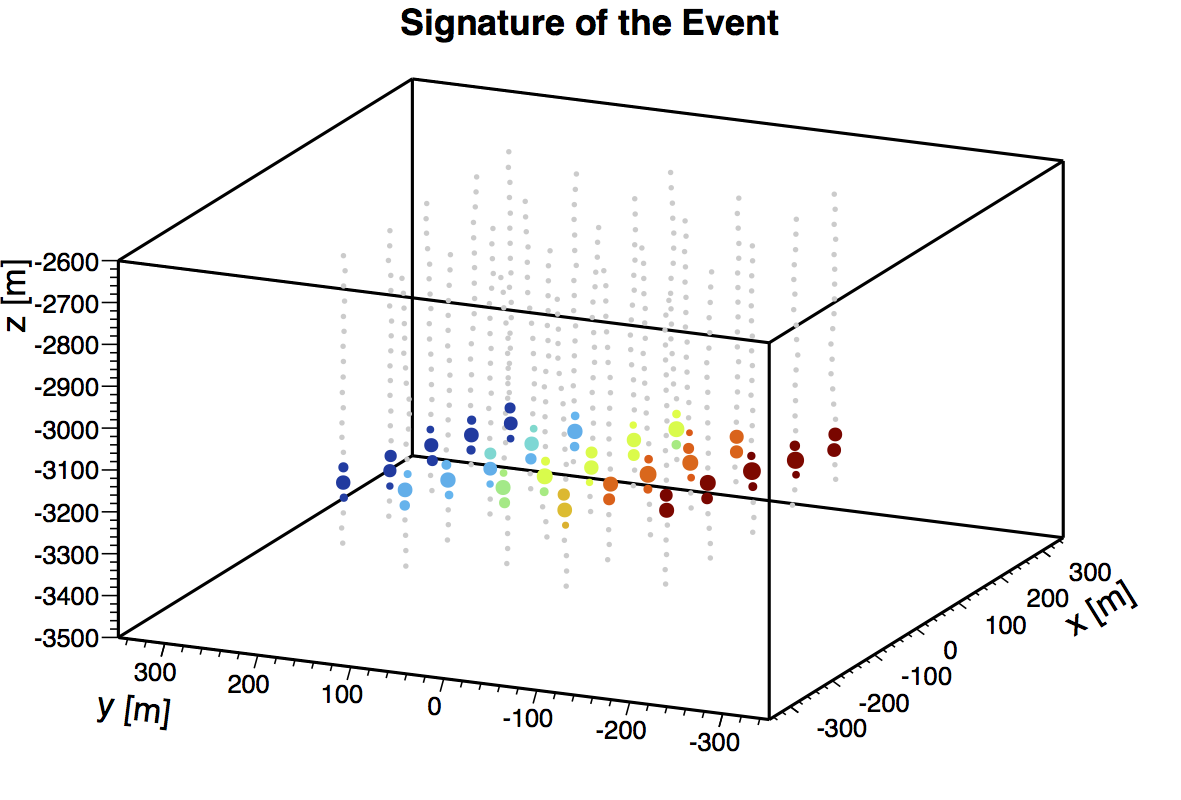}
  \caption{The signature of a neutrino event, simulated in a KM3NeT-like acoustic detector with 25 lines spaced 90\,m apart. Each dot represents an optical module with an integrated acoustic sensor, while the color indicates the arrival time of the acoustic signal (red: early, blue: late, grey: not triggered). The size of the dot represents the signal amplitude. The neutrino interacted at a distance of 1.8\,km from the detector center with an energy of $10^{21}$\,eV.}
  \label{fig:signature}
\end{figure}

The experience from the data taken with the AMADEUS set-up has shown that several stages of signal classification are needed for background suppression. For the AMADEUS data acquisition, machine learning algorithms are used to identify bipolar pulses. They further discard clustered events, as neutrinos are not expected to produce several signals from one position within a short period of time. The remaining event density after these cuts is still approximately $100\;\mathrm{events/km^{3}/yr}$~\cite{ref9}. These events cannot be caused by neutrino interactions, as this number is orders of magnitude bigger than any model would predict. %
% Therefore, an additional method to classify the signal has to be found. There is so far unused information available from the signal, which is the sound emission pattern (the ``pancake plane'' for neutrinos). 
The sound emission pattern (the ``pancake plane'' for neutrinos) provides a so far unused additional signal information. Utilizing this information is not possible with the AMADEUS set-up, since it requires a 3 dimensional detector while the available set-up consists of only six sensor clusters arranged in a vertical plane. However, the first phase of the future neutrino telescope KM3NeT will consist of 31 lines with 18 acoustic sensors per line. They will be split in two blocks, a large one with 24 and a smaller one with seven lines. Only the large block is considered here in order to test the possibility of the classification by means of the emission geometry. A simulated neutrino signal for this configuration is shown in Fig.~\ref{fig:signature}. The simulation of the signals was performed with the simulation chain described in the previous section. A background sample was also created with signals from the acoustic positioning system, spherical sound signals from random positions and completely random signatures. 

The classification is carried out in the following order: first, the vertex of the event is reconstructed. Then, a singular value decomposition (SVD) of a matrix with the positions of the hydrophones that triggered the signal is performed. The SVD also reconstructs the plane within the detector, since the singular vector belonging to the smallest singular value is also the normal vector of that plane. The resulting plane is then used to calculate other parameters, e.g. the distance of the reconstructed vertex to the plane, the average distance of the hits from the plane, the center of gravity of the hits and the time residual if the reconstructed vertex is projected into the plane. All of these features are normalized to eliminate any dependency on the detector geometry. 

All values are stored in a feature vector and the classification is done by machine learning algorithms implemented in the OpenCV library~\cite{ref10}. Several different methods were tested: decision trees, random trees, boosted trees and support vector machines. The best performance is achieved with boosted trees, with a recognition rate of 98.8\,\% for neutrino signals and 99.7\,\% for the background events. The training of the algorithm was checked over a range of sample sizes from 100 to 45\,000 events (with 50\,\% neutrinos), showing no improvement for sizes exceeding 10\,000 events. The test data set always comprised 20\,000 events with 5\,000 neutrinos. This model, trained for the first phase of the KM3NeT telescope (24 lines, 18 sensors/line), was also tested with a similar set-up with fewer lines, loosing only about 2\,\% efficiency in the recognition rates. The algorithm will be trained for a specific detector for its future application. However this shows that the approach of using only normalized values could handle a change of the detector, e.g. a failure of a line, without any problems. First tests with ray traced sound signals additionally indicate that the classification will also work here, but the recognition rates of neutrino signals drop below 90\,\% unless the algorithm is trained with a new data set. However, using a model trained with data from the ray tracing sound propagation to classify data simulated with straight sound propagation resulted in no change of the recognition rates.

The application to real data would require additional testing. The events simulated here are pure, meaning that it is not possible that a hit from another event is falsely added to it. So, the influence of falsely identified hits on the performance of the classification has to be tested as well as if a second stage of hit selection is needed after the vertex reconstruction, e.g. using a random sample consensus algorithm, and how it affects the recognition rates.

\section{Summary and Outlook}
A fully operational simulation chain for acoustic particle detection written in C++ is available within the software framework SeaTray, only requiring specific hardware features like the system response and inherent noise to adapt to a new detector set up. It is being used to develop a signal classification for neutrinos and to estimate the performance of future detectors with the size of the KM3NeT neutrino telescope.

This work is supported by the German Government (BMBF) with grants 05A08WE1 and 05A11WE1. 

%%%%%%%%%%%%%%%%%%%%%%%%%%%%%%%%%%%%%%%%%%%%%%%%
%% BACKMATTER
%%%%%%%%%%%%%%%%%%%%%%%%%%%%%%%%%%%%%%%%%%%%%%%%

%\begin{theacknowledgments}
%  Infandum, regina, iubes renovare dolorem, Troianas ut opes et
%  lamentabile regnum cruerint Danai; quaeque ipse miserrima vidi, et
%  quorum pars magna fui. Quis talia fando Myrmidonum Dolopumve aut duri
%  miles Ulixi temperet a lacrimis?
%\end{theacknowledgments}



\bibliographystyle{aipproc}   % if natbib is available

\bibliography{lit_ac_sim}

\IfFileExists{\jobname.bbl}{}
 {\typeout{}
  \typeout{******************************************}
  \typeout{** Please run "bibtex \jobname" to optain}
  \typeout{** the bibliography and then re-run LaTeX}
  \typeout{** twice to fix the references!}
  \typeout{******************************************}
  \typeout{}
 }

\end{document}